\documentclass[12pt,preprint]{aastex}

\shorttitle{IRAS 04000+5052}
\shortauthors{Esteban et al.}

\begin{document}

\title{IRAS 04000+5052: A not so compact, not so metal-poor \ion{H}{2} region
\footnote{Based on observations made with telescopes operated on 
the island of La Palma by the Isaac Newton Group of Telescopes and
Nordic Optical Telescope in the 
Spanish Observatory of Roque de Los Muchachos of the Instituto de 
Astrof\'\i sica de Canarias.}}

\author{C\'esar Esteban}
\author{Luis L\'opez-Mart\'\i n}
\author{\'Angel R. L\'opez-S\'anchez} 
\author{Bernab\'e Cedr\'es}
\and
\author{Jorge Garc\'\i a-Rojas}

\affil{Instituto de Astrof\'\i sica de Canarias, c/V\'\i a L\'actea s/n, 38200, La Laguna, Tenerife, Spain}

\email{cel@iac.es, luislm@iac.es, angelrls@iac.es, jogarcia@iac.es, bce@iac.es}

\begin{abstract}

We present new observations of IRAS 04000+5052, a Galactic \ion{H}{2} region associated with 
a young stellar cluster and possibly located at the Perseus arm. Contrary to previous claims, we have found that 
this object is not a compact metal-poor \ion{H}{2} region. The electron density and chemical composition  
of the nebula are similar to those of normal \ion{H}{2} regions of the Galactic disk. The radial velocity of the 
ionized gas coincides with that obtained from CO observations, indicating that the nebula is associated with a 
molecular cloud. Probably, the ionizing source of the \ion{H}{2} region is a Herbig Be star of B0.5 spectral type. 

\end{abstract}


\section{Introduction}

IRAS 04000+5052 (Galactic coordinates: 150.86, -1.12) has an IRAS LRS 
spectrum and IRAS fluxes consistent with an \ion{H}{2} region \citep{wan93}. 
\cite{wou89} found $^{12}$CO emission in the direction of 
the object and calculate galactocentric and heliocentric distances of 12.04 
and 3.88 kpc, respectively (assuming a solar galactocentric distance of 8.5 kpc) 
for the molecular emission. 
This indicates that IRAS 04000+5052 --if really associated with the molecular cloud-- 
is located beyond the solar circle, and 
beyond the Perseus arm. \cite{wan93} found a near-infrared (NIR) 
variable source associated with the IRAS source, probably a young object 
embedded in the \ion{H}{2} region. In a more recent study, Wang, Wei, \& Hu (2002) 
have carried out a more complete study of the object, based 
on NIR imagery and optical spectroscopy. These authors find 
an isolated high-mass protostellar infrared cluster of 15 objects associated 
with the IRAS source, where a B0.5-type star (their IRS~7 object) is the ionizing source of the \ion{H}{2} 
region. \cite{wan02} classify IRAS 04000+5052 as a compact \ion{H}{2} region 
based on the criteria of its size and location in a star-forming 
molecular cloud. These authors also estimate the [\ion{N}{2}]/H$\alpha$ ratio of 
the nebula from Gaussian fitting of the blend of H$\alpha$ and [\ion{N}{2}] 6548 and 
6584 \AA\ lines measured in their low-resolution spectrum ($\sim$ 10 \AA). 
The [\ion{N}{2}]/H$\alpha$ ratio they find is about 1/16.3, which is the 
lowest value ever reported for a Galactic \ion{H}{2} region; this result suggests  
that IRAS 04000+5052 is a rather peculiar metal-poor nebula and merits 
further investigation.

We have carried out optical spectroscopy and imagery of IRAS 04000+5052 in 
order to obtain a more precise determination of its 
chemical composition and to disentangle the true nature of the object.

\section{Observations}

Intermediate-resolution spectroscopy was obtained in 2002 December 27, with the ISIS Spectrograph  
at the 4.2 m William Herschel Telescope (WHT) of Roque de los Muchachos Observatory 
in La Palma, Spain. Two different CCDs were used at the blue and red arms of the spectrograph: 
an EEV CCD with a configuration 4096 $\times$ 2048 pixels of 13.5 $\mu$m in the blue arm and a 
Marconi CCD with 4700 $\times$ 2148 pixels of 13.5 $\mu$m in the red arm. The dichroic used to separate 
the blue and red beams was set at 5400 \AA. The slit was 3.7$'$ long and 1.03$''$ wide. Two gratings 
were used, the R1200B in the blue arm and the R316R in the red arm. These gratings give reciprocal 
dispersions of 17 and 62 \AA\ mm$^{-1}$, and effective spectral resolutions of 0.86 and 3.81 \AA\ 
for the blue and red arms, respectively. The blue spectra cover from 4456 to 5484 \AA\ and the red 
ones from 5370 to 8690 \AA. The spatial scale is 0.20$''$ pixel$^{-1}$ in both arms. 

The slit was centered on the brightest part of the nebula with a position angle of 90$^\circ$ 
(east-west). Three 600 seconds exposures were taken and combined to obtain good signal-to-noise and an 
appropriate removal of cosmic rays. The spectra were wavelength calibrated with a CuNe+CuAr lamp. 
The correction for atmospheric extinction was performed using the average curve for continuous 
atmospheric extinction at Roque de los Muchachos Observatory. The absolute flux calibration of the 
spectra was achieved by observations of the standard stars Feige~15, Feige~110, H~600 and Hz~44. 
All the CCD frames were reduced using the standard $IRAF$\footnote{IRAF is distributed by NOAO, which is
operated by AURA, under cooperative agreement with NSF} TWODSPEC reduction package to perform bias 
correction, flat-fielding, cosmic-ray rejection, wavelength and flux calibration, and sky subtraction. 

The optical CCD images were obtained in 2004 January 21
at the 2.56 m Nordic Optical Telescope (NOT) at Roque de los Muchachos Observatory (La Palma, 
Spain). We used the ALFOSC (Andalucia Faint Object Spectrograph and Camera) in image mode with a 
E2V CCD detector (2048 $\times$ 2048 pixels) with a pixel size of 13.5 $\mu$m and spatial resolution 
of 0.19$''$ pixel$^{-1}$. The narrow-band filters for 
the H$\alpha$ and continuum have a full width at half maximum (FWHM) of 33 and 51 \AA, respectively. The continuum filter has its  
central wavelength at 6891 \AA. The average seeing of the images was 1.2$''$. Four exposures of 300 seconds in 
H$\alpha$ and 3 exposures of 150 seconds 
in the continuum were added to obtain a good signal-to-noise ratio and an appropriate removal of cosmic 
rays in the final images. The images were bias-subtracted, flat-fielded, and flux calibrated following 
standard procedures making use of $IRAF$ software, in particular the FOCAS package. The absolute flux calibration was 
performed using the spectrophotometric standard stars Feige 110 and HR 1544, and following the method proposed by 
\cite{bar94}. 
Figure \ref{fig1} shows the resulting H$\alpha$ image of IRAS~04000+5052.

\section{Results and discussion}

\begin{figure}
\resizebox{\hsize}{!}{\includegraphics{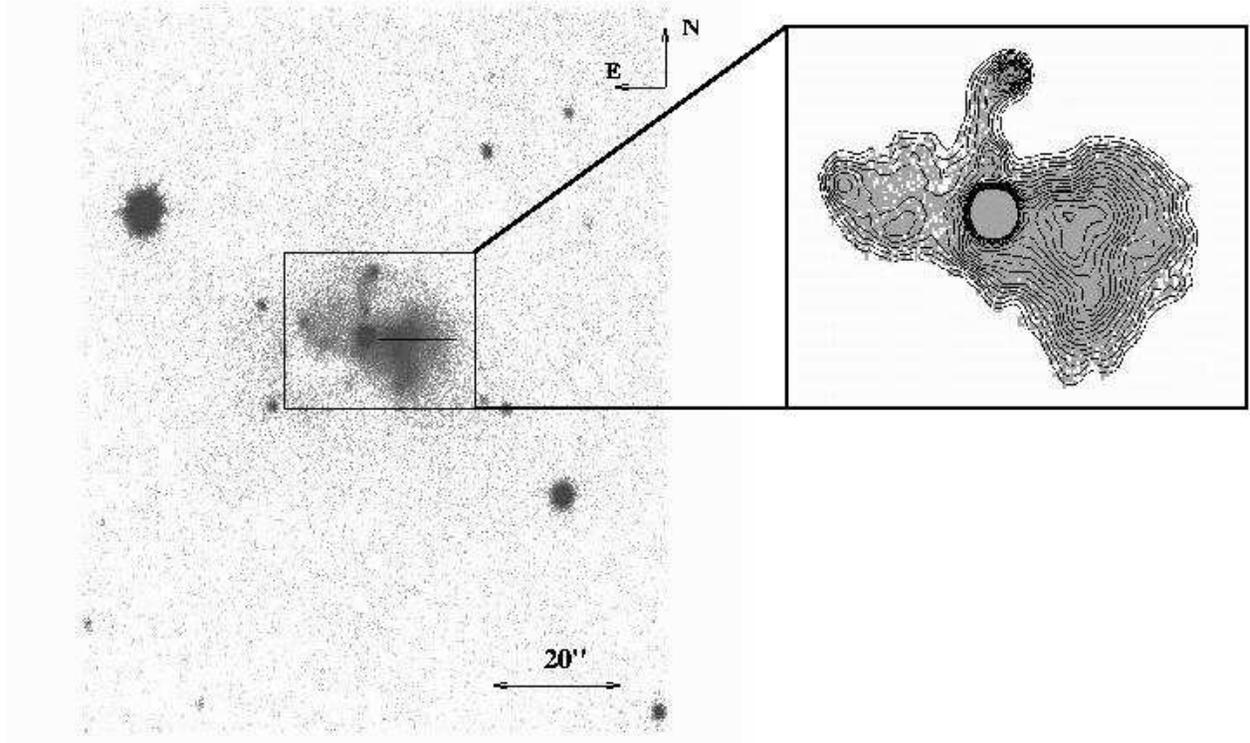}}
\caption{CCD H$\alpha$ image of IRAS 04000+5052. The position and spatial extension of the zone from 
which the one-dimensional spectra have been extracted is indicated. The box on the right shows an enlargement of the central 
part of the continuum-subtracted image of the nebula in gray scale with contours.\label{fig1}}
\end{figure}

The H$\alpha$ image of the nebula (Figure \ref{fig1}) shows much more detail than the broad-band NIR and Br$\gamma$ 
images published by \cite{wan02}. IRAS~04000+5052 has a patchy structure with the presence of dark zones and a very faint extended 
ionized halo. The morphology is typical of an \ion{H}{2} region associated with a substantial amount of obscuring 
material and a star-forming zone. 

\begin{figure}
\plotone{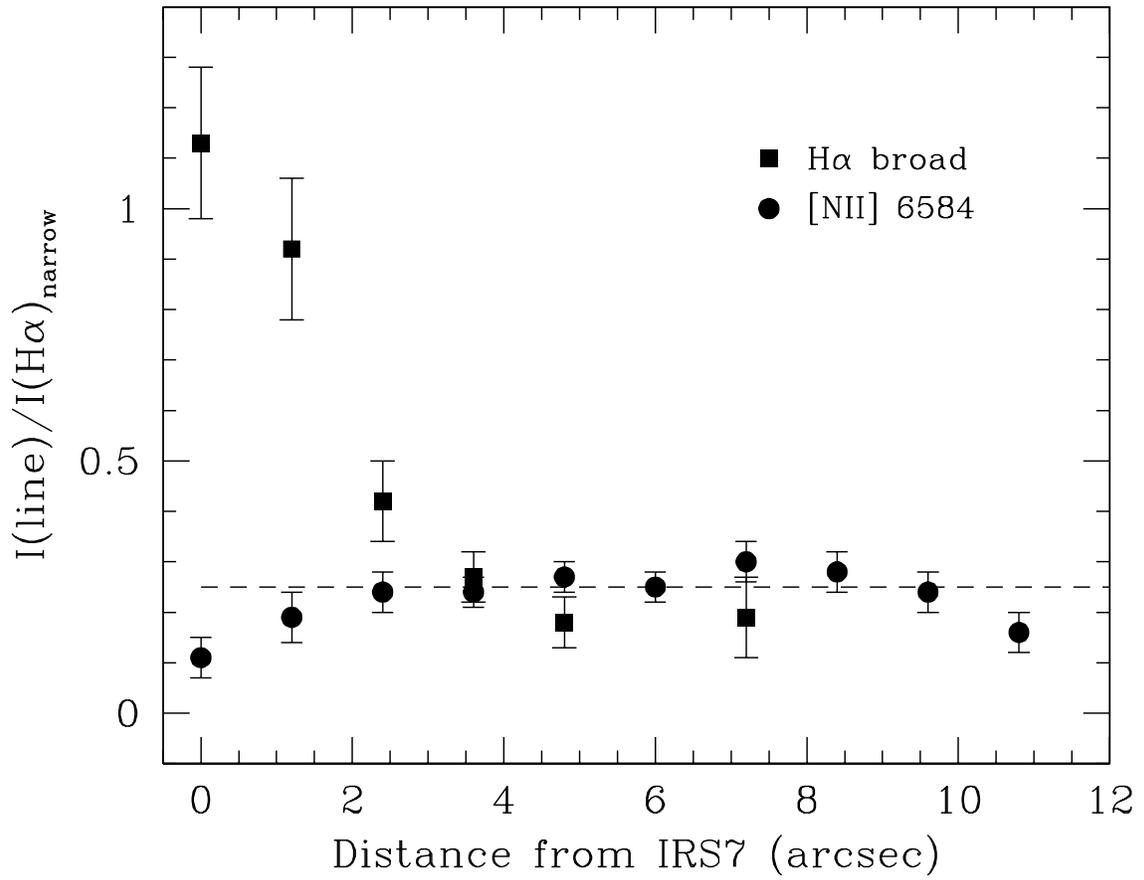}
\caption{Emission line ratios of the broad component of H$\alpha$ and nebular [\ion{N}{2}] with respect to nebular --narrow-- 
 H$\alpha$ emission along part of the slit (from the ionizing star IRS~7 to the west). The spatial increments 
 are 1.2$''$-long. The contribution of the
 broad H$\alpha$ emission is more important near IRS~7 while the [\ion{N}{2}] emission is quite constant 
 along the slit. \label{fig2}}
\end{figure}

\begin{figure}
\resizebox{\hsize}{!}{\includegraphics[angle=90]{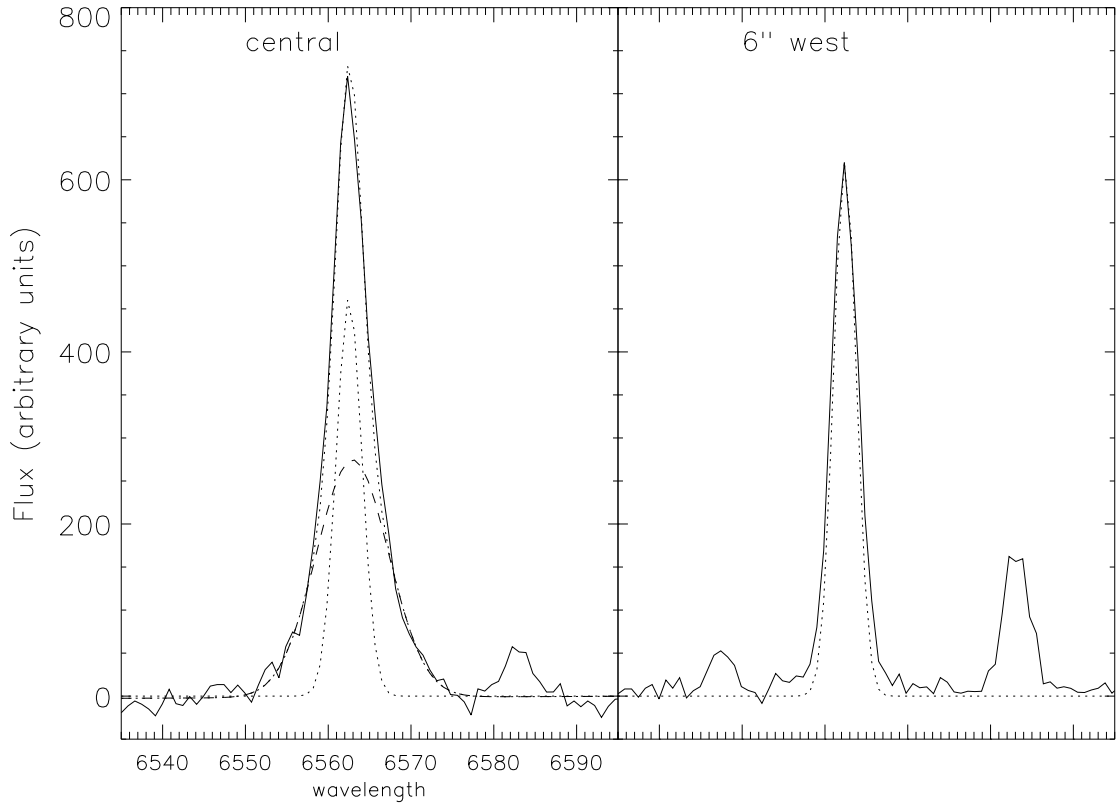}}
\caption{Spectra of a 1.2$''$-long slice passing through the central star of IRAS 04000+5052 \ion{H}{2} region (left) and an 
offcenter region 6 arcsec to the west of the central star (right). The Gaussian fits to the line profiles --individual and combined-- 
are included as discontinuous curves. The FWHM and nature of the components are given and discussed in the text. 
The flux is in arbitrary units and the scaling factor used in both boxes is different. \label{fig3}}
\end{figure}

The bidimensional spectrum shows that H$\alpha$ is the brightest line of the nebula. The profile of this line is rather complex 
along a substantial part of the spatial extension of the nebula. This profile can be reproduced by the combination of two 
Gaussian components of different widths. The relative contribution of the broad component is more important at the position 
of the stellar continuum of IRS~7 (the variable NIR source found by Wang et al. 2002 that corresponds to  
the brightest star embedded in the nebula) and decreases as one goes farther from the star. 
This fact indicates that the broad component is produced by the light of a stellar emission feature scattered by the dust 
mixed with the ionized gas. In Figure \ref{fig2}, we show the relative contribution of the broad and narrow components of 
the H$\alpha$ line (as well as the [\ion{N}{2}]/H$\alpha$ intensity ratio) for different slices of the spectrum 
along the brightest part of the nebula. 
The line profile was analyzed via a double Gaussian fitting using Starlink DIPSO software
package \citep{how90}. For each Gaussian fit, 
DIPSO gives the fit parameters (radial velocity centroid, 
integrated flux, FWHM, etc) and their associated statistical errors. In Figure \ref{fig3} we show examples of Gaussian fits 
in two small different zones of the nebula. The broad (stellar) component is specially important in the zone surrounding IRS~7 
(Figure \ref{fig3}, left; and also Figure \ref{fig2}). 
The intrinsic (corrected for instrumental profile) FWHM of the broad component is about 
520 km s$^{-1}$ in that zone. In contrast, the intrinsic FWHM of the narrow (nebular) component of 
H$\alpha$ is typically of the order of 
30$-$50 km s$^{-1}$. In the right panel of Figure \ref{fig3} we show an example of a zone without a noticeable contribution 
of the broad component. The broad H$\alpha$ component should be related to the reflection of a stellar emission 
feature produced by a circumstellar disk or strong stellar 
winds associated with IRS~7, the most probable ionization source of the \ion{H}{2} region.

\begin{figure}
\plotone{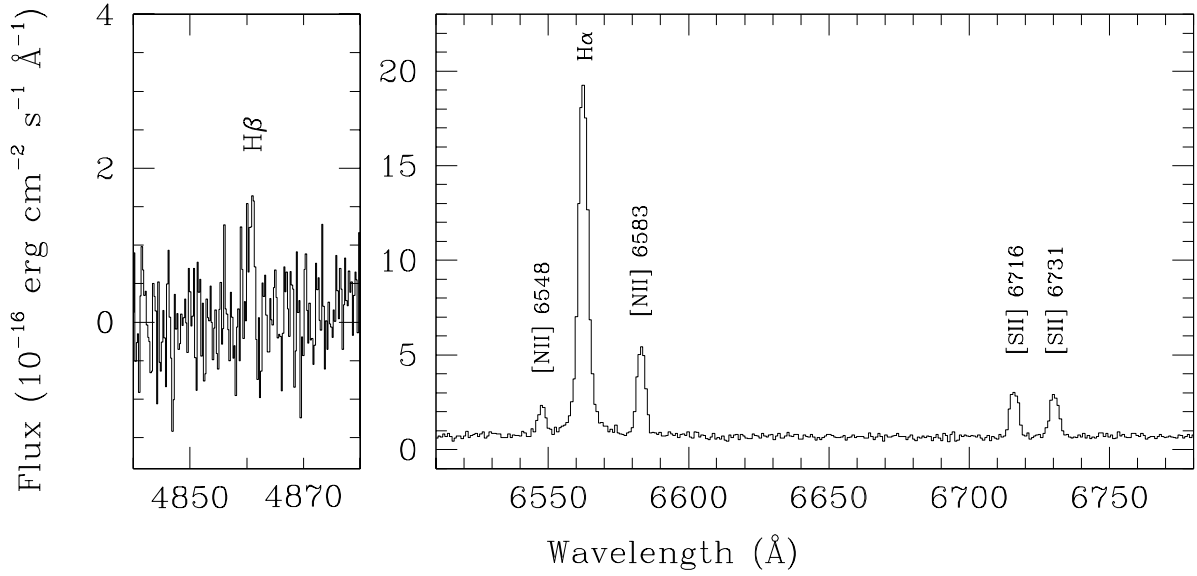}
\caption{Parts of the blue (left) and red (right) spectra of IRAS 04000+5052 showing all the emission lines measured in the nebula. 
H$\alpha$ and the [\ion{N}{2}] doublet are clearly 
resolved (see Figure 2 of Wang et al. 2002 for comparison).\label{fig4}}
\end{figure}

We extracted a single onedimensional spectrum integrating from 2 to 15 arcsec to the west 
of IRS~7 (shown in Figure 1) in order to estimate the physical conditions and chemical 
abundances of the ionized gas. We avoided the zone around IRS~7 because the important contribution of the broad component 
to the total H$\alpha$ flux. Figure \ref{fig4} shows parts of the blue and red spectra of IRAS~04000+5052. H$\beta$ 
is the only line detected in the blue spectrum and shows a rather low signal-to-noise ratio. Line intensities were 
measured integrating all the flux between two given limits and over a local continuum estimated by eye. 
In Table 1 we include the observed and 
derredened line intensity ratios with respect to $I$(H$\beta$) of the emission lines detected in the integrated onedimensional 
spectrum. A relatively small contribution of the broad component of H$\alpha$ is still present in this spectrum. 
In Table 1 we include the individual parameters of the two components of H$\alpha$. 
The observed line intensities have been corrected for interstellar reddening using the \cite{whi58} law and the 
reddening constant, $C(H\beta)$. This constant has been derived from the ratio between the observed H$\beta$ 
and H$\alpha$ --narrow component-- line intensities as compared with the theoretical values expected for case B recombination 
using \cite{bro71} and assuming an electron temperature of 10,000 K. The value of $C(H\beta)$ is 3.1$\pm$0.5, 
which corresponds to an extinction of $A_V$ = 6.5$\pm$1.1 magnitudes. \cite{wan02} estimate a lower limit of 
$A_V$ of about 2.64 magnitudes from their upper limit to the intensity of H$\beta$, which is not detected in their 
spectrum. In Table 1, we have also included the velocity (with respect to the local standard of rest, $v_{LSR}$) corresponding 
to the centroid of the single or double Gaussian fitting of the lines. The reddening-corrected H$\beta$ line flux 
of the onedimensional spectrum is 3${\pm 1}$ $\times$ 10$^{-13}$ erg cm$^{-2}$ s$^{-1}$.
  
The weighted mean of the $v_{LSR}$ of the different lines --except that of the broad H$\alpha$ component-- is 
$-$30.1$\pm$0.5, which is entirely consistent with the velocity obtained by \cite{wou89} from the $^{12}$CO($J$ =
1$-$0) emission in the direction to the optical nebula, $v_{LSR}$ = $-$30.5$\pm$0.2 km s$^{-1}$. This result proves 
that the nebula is certainly associated with the molecular emission reported by \cite{wou89}. These authors determined 
a kinematic heliocentric 
distance of 3.88 kpc for IRAS~04000+5052. On the other hand, \cite{wan02} derived a photometric distance of 4.29 kpc 
using the observed $K$ magnitude of IRS~7, its estimated B0.5 spectral type, and their lower limit of the 
extinction ($A_K$ = 0.11 $\times$ $A_V$ = 0.29 mag). Following the same procedure of those authors, we have recalculated 
the photometric distance of IRAS~04000+5052 
using our more precise value of $A_K$ = 0.72 mag, finding a distance of 3.55 kpc, which is also fairly similar to 
the kinematic one. However, those authors implicitly assume a $M_K$ $\approx$ $-$3.85 mag for their 
calculations without quoting any reference. We have revised the recent literature finding that the absolute $K$ magnitude given 
by \cite{wan02} seems to be excessively bright. Hanson, Howarth, \& Conti (1997) obtain $M_K$ = $-$1.56 mag for a ZAMS 
B0.5 V star. 
On the other hand, \cite{weg00} finds an average $M_V$ = $-$2.75 for a sample of 12 Galactic B0.5 IV$-$V stars with $Hipparcos$ 
parallaxes. Using the intrinsic $V-K$ color of B0.5 V stars of $-$0.79 obtained by \cite{kor83} we derive an expected 
$M_K$ = $-$1.96 for this kind of stars. These two determinations of the absolute $K$ magnitude are about two magnitudes 
fainter than the value adopted by \cite{wan02}. \cite{weg00} indicates that it is common that Be stars are brighter than normal B stars 
of the corresponding spectral types, about 0.5 mag in the case of B0.5 stars. As we can see, the appropriate value 
of the absolute magnitude of the ionizing star is uncertain and this makes difficult to fix a precise value of the photometric distance for the 
object. If we assume a $M_K$ between $-$1.5 and $-$2.5 mag, this would imply a heliocentric distance between 1.2 and 1.9 kpc, and 
that the object could belong to the Perseus arm (located at about 2 kpc, see Russeil 2003). This result is contrary to the claim of 
\cite{wan02} that IRAS~04000+5052 is located beyond that spiral arm. Differences between kinematic and photometric 
distances are rather usual, in fact, \cite{rus03} in its recent revision of the distribution of star-forming complexes and the 
structure of the Galaxy, indicates that velocity anomalies are common in the Perseus arm. 

\cite{wan02} estimate the number of Lyman continuum photons, $N$(Lyc), that ionizes the nebula considering the radio 
continuum flux density at 1.4 GHz obtained from the NRAO VLA Sky Survey (NVSS; Condon et al. 1998) and assuming a heliocentric 
distance of 3.88 kpc, finding 
log $N$(Lyc) = 46.27, value that corresponds to a ZAMS star of spectral type B0.5 \citep{pan73}. We have 
obtained the integrated flux from our dereddened flux calibrated and continuum-subtracted H$\alpha$ CCD image of the nebula 
finding a value of 6.64$^{+2.34}_{-1.98}$  10$^{-11}$ erg cm$^{-2}$ s$^{-1}$. Assuming a representative value of the heliocentric 
distance of about 1.5 kpc (we have not considered errors for the distance in the following because it is rather undefined), 
we obtain an 
H$\alpha$ luminosity of $L_{\rm H\alpha}$ = 1.79$^{+0.65}_{-0.54} \times 10^{34}$ erg s$^{-1}$, which 
corresponds --assuming an ionization-bounded nebula-- to log $N$(Lyc) = 46.12$^{+0.13}_{-0.16}$ (i. e. Osterbrock 1989). This 
ionizing flux is consistent with a ZAMS star of B0.5 spectral type, in --accidental-- agreement with the results of \citet{wan02}. 
In addition, we consider that the central object IRS~7 
can be classified as a Herbig Be star attending to several facts: a) its early B spectral type; 
b) the presence of broad H$\alpha$ emission related to the star; c) its pre-main sequence nature (due to its association with a 
very young stellar cluster, an \ion{H}{2} region, and a molecular cloud). The corresponding galactocentric distance of the 
object is about 9.8 kpc (assuming the Sun at 8.5 kpc).

\begin{figure}
\resizebox{\hsize}{!}{\includegraphics{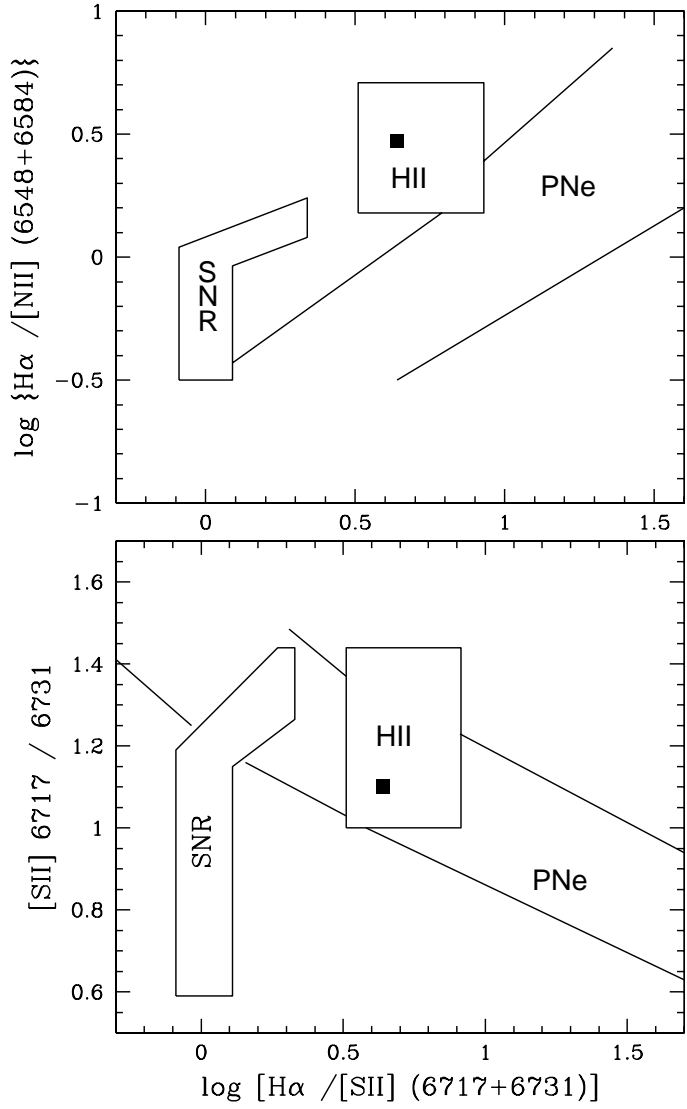}}
\caption{Diagnostic diagrams showing the emission line ratios of IRAS 04000+5052 indicating 
the \ion{H}{2} region nature of the object (adapted from Sabbadin et al. 1977). \label{fig5}}
\end{figure}

From Figure \ref{fig4} and Table 1 it is clear that H$\alpha$ and the two lines of [\ion{N}{2}] doublet are well resolved 
in our spectrum, 
and that the [\ion{N}{2}](6584 \AA)/H$\alpha$ ratio is 0.25, much larger than the value of 0.06 derived by \cite{wan02} 
from a Gaussian fitting of an unresolved blend of those lines. This result refutes the statement by \cite{wan02} 
that IRAS~04000+5052 is a metal-poor nebula with the lowest [\ion{N}{2}]/H$\alpha$ line ratio known 
in Galactic \ion{H}{2} regions. In contrast, our value of 0.25 coincides with the average of the 
[\ion{N}{2}](6584 \AA)/H$\alpha$ ratio observed in a sample of 15 Galactic \ion{H}{2} regions \citep{haw78}. 
On the other hand, the [\ion{N}{2}]/H$\alpha$ ratio --valuated only for the nebular or narrow component of H$\alpha$-- 
is almost constant along the nebula, as it can be seen in Figure \ref{fig2}. 
We cannot determine the electron temperature of the ionized gas and, therefore, it is not possible to obtain a 
direct estimation of the chemical abundances of the nebula. In any case, we can indirectly estimate 
the metallicity making use of empirical calibrations. Denicol\'o, Terlevich, \& Terlevich (2002) propose a 
calibration of the O/H ratio vs. the 
[\ion{N}{2}](6584 \AA)/H$\alpha$ ratio. Using this relation we obtain a oxygen abundance of 
12 + log(O/H) $\sim$ 8.70, a value typical of \ion{H}{2} regions of the solar vicinity. 
In Figure \ref{fig5}, we show two diagnostic 
diagrams (adapted from Sabbadin, Minello, \& Bianchini 1977) involving the locus of several line ratios we have observed in the nebula. 
These diagrams clearly indicate that IRAS~04000+5052 is radiatively excited and has a spectrum of a typical 
\ion{H}{2} region, consistently with that suggested by its IRAS LRS spectrum, IRAS fluxes, and morphology. 

The detection and measurement of the [\ion{S}{2}] doublet at 6716 and 6731 \AA\ allows to deriving the electron density 
of the nebula. Using the five-level program for the analysis of emission-line nebulae included in IRAF 
NEBULAR task \citep{sha95} and our observed [\ion{S}{2}] 6717/6731 ratio we obtain an electron density of 
540$^{+400}_{-200}$ cm$^{-3}$. This is a density similar to that observed from collisionally excited line ratios 
in other normal Galactic \ion{H}{2} regions. 
For example, the central parts of the Orion nebula have electron densities between 3000 and 5000 cm$^{-3}$ 
\citep{est98} and M17 between 500 and 1000 cm$^{-3}$ \citep{est99}. \cite{vil96} obtain 
densities between $\leq$100 cm$^{-3}$ and 550 cm$^{-3}$ for a sample of \ion{H}{2} regions located toward the Galactic 
anticenter. From Figure \ref{fig1} and the adopted distance of 1.5 kpc it is possible to estimate the 
approximate linear dimensions of the nebula, which are about 0.19 pc $\times$ 0.14 pc. 

From our dereddened value of the H$\alpha$ flux and the diameter of the nebula we can estimate its root mean 
square density using the usual formulae (i.e. Gurzadyan 1997), which is $N_e$(rms) = 950$\pm$150 cm$^{-3}$. 
This value is larger than the density obtained from the [\ion{S}{2}] doublet but consistent within the uncertainties, 
result that indicates that the volume filling factor of the nebula should be about 1. 
It is necessary to remark that the exact value of $N_e$(rms) is extremely dependent on the assumed distance, which 
is rather uncertain in this case. The corresponding ionized mass of the nebula is very small: 0.11$\pm$0.04 M$_\odot$. 
In Table 2 we 
compare some physical parameters of IRAS~04000+5052 with those typical of compact and "classical" --normal-- \ion{H}{2} 
regions \citep{kur02}. \cite{wan02} propose that IRAS~04000+5052 
is a compact \ion{H}{2} region attending to size considerations and its association with a molecular cloud but lacking 
of any information about electron density. From Table 2 we can see that, since the size and mass of the nebula are 
of the order of those of typical compact \ion{H}{2} regions, the density is consistent with the typical values of normal \ion{H}{2} 
regions. We consider that 
the small linear size an ionized mass of the nebula is due to the rather weak ionizing flux of the star, which is relatively 
cold because of its estimated B0.5 spectral type. Therefore,  
we consider that IRAS~04000+5052 should be considered a relatively normal small \ion{H}{2} region ionized 
by a rather weak and cold ionizing source.

\section{Conclusions}

We present new observations of IRAS~04000+5052, a faint Galactic \ion{H}{2} region associated with a small and isolated 
protostellar cluster probably located at the Perseus arm. We have found that this \ion{H}{2} region is associated with a molecular
cloud and that it cannot be considered 
a compact metal-poor \ion{H}{2} region located beyond the Perseus arm as it was proposed by \cite{wan02}. It is ionized by a stellar source of B0.5 
spectral type, perhaps a Herbig Be star. Its corresponding 
ionizing flux is rather weak and the size of its associated H$^+$ sphere is correspondingly small, with a diameter of 
the order of 0.2 pc. 
The \ion{H}{2} region is very faint, highly reddened and patchy, as expected due to its association with a molecular cloud and a 
zone of recent star formation. 
The electron density of the nebula is consistent with that of typical 
\ion{H}{2} regions and not of compact ones. Definitively, the nebula is not a metal--poor object, 
its [\ion{N}{2}](6584 \AA)/H$\alpha$ ratio is 0.25, larger 
than the value of 0.06 previously determined by \cite{wan02} and consistent with that of normal \ion{H}{2} regions of the 
Galactic disk. 

\acknowledgments
We are grateful to an anonymous referee for his/her patience and very useful comments that helped to improve the paper.

\begin{table*}
\centering
  \caption{Line intensity ratios with respect to I(H$\beta$) = 100}
  \label{table1}  
  \footnotesize
  \begin{tabular}{lrccc}
    \hline\hline
	\noalign{\smallskip}
	& & & & $v_{LSR}$ \\
    Line & $f$($\lambda$)& $F$($\lambda$)/$F$(H$\beta$) &  $I$($\lambda$)/$I$(H$\beta$) & (km s$^{-1}$) \\
	\hline    
	\noalign{\smallskip}
 4861 H$\beta$ & 0.00 & 100$\pm$35& 100$\pm$35& $-$46$\pm$22 \\
 6548 $[$\ion{N}{2}$]$ & $-$0.34 & 246$\pm$25 & 22$\pm$2 & $-$30$\pm$5 \\
 6563 H$\alpha\rm^a$ & $-$0.34 & 3190$\pm$65 & 289$\pm$6 & $-$30.1$\pm$0.6 \\
 6563 H$\alpha\rm^b$ & $-$0.34 & 323$\pm$32 & 29$\pm$3 & $-$8$\pm$16 \\ 
 6584 $[$\ion{N}{2}$]$ & $-$0.34 & 794$\pm$40 & 72$\pm$4 & $-$26$\pm$2 \\
 6716 $[$\ion{S}{2}$]$ & $-$0.36 & 410$\pm$28 & 32$\pm$2 & $-$32$\pm$2 \\
 6731 $[$\ion{S}{2}$]$ & $-$0.36 & 402$\pm$28 & 32$\pm$2 & $-$34$\pm$3 \\
 \noalign{\smallskip}
    \hline\hline
  \end{tabular}
\begin{flushleft}
  $\rm^a$ Nebular (narrow) component \\
  $\rm^b$ Contribution of dust scattered stellar H$\alpha$ emission feature\\
  \end{flushleft}
\end{table*}

\begin{table*}
\centering
  \caption{Comparison of physical parameters}
  \label{table2}  
  \footnotesize
  \begin{tabular}{lcccc}
    \hline\hline
	\noalign{\smallskip}
	Class of & Size & $N_e$ & EM & Ionized Mass \\
    \ion{H}{2} Region & (pc) &  (cm$^{-3}$) & (pc cm$^{-6}$) & (M$_\odot$) \\
	\hline    
	\noalign{\smallskip}
 Compact$\rm^a$ & $\lesssim$ 0.5 & $\gtrsim$ 5 $\times$ 10$^3$ & $\gtrsim$ 10$^7$ & $\sim$ 1 \\
 Classical$\rm^a$ & $\sim$ 10 & $\sim$ 100 & $\sim$ 100 & $\sim$ 10$^5$ \\
 \noalign{\smallskip}
 IRAS~04000+5052 & $\sim$ 0.2 & 540 $^{+400}_{-200}$ & $\sim$ 5.8 $\times$ 10$^4$ & 0.11 $\pm$ 0.04 \\
 \noalign{\smallskip}
    \hline\hline
  \end{tabular}
\begin{flushleft}
  $\rm^a$ Reference values taken from Kurtz \& Franco (2002) \\
\end{flushleft}
\end{table*}

\clearpage
\clearpage
\clearpage
\clearpage
\clearpage

\end{document}